\documentclass{article}
\usepackage{graphicx}
\begin{document}
\title{The cosmic censorship conjecture in a higher dimensional spacetime with an interacting vacuum energy }
\author{M. Campos\\
              Physics Department, Roraima Federal University \\
              Tel.: +55-95-36213139\\
              Fax: +55-95-36213137\\
              {miguel.campos@ufrr.br}   }        

\date{Received: date / Accepted: date}
\maketitle
\begin{abstract}
Presently, the inclusion of the vacuum energy in the energy momentum tensor, and the inclusion of the extra dimensions in the spacetime, can not be rule out of the research in gravitation.  In this work we study the influence of the vacuum energy in the collapse process of a stellar fluid, and consequently for the cosmic censorship conjecture, considering a homogeneous and isotropic spacetime with arbitrary number of dimensions. We discuss the active gravitational mass of the black hole formed, where the vacuum energy and the number of dimensions has a crucial role in the process.
\end{abstract}

\section{Introduction}
\label{intro}
There are several possibilities in the scientific literature to explain the accelerated expansion of the universe, that was detected recently making use of different cosmological experiments, but the concordance of the $\Lambda$CDM
 model with the observational data, materialized this model as paradigmatic in the present time.  On the other hand, the most theoretically appealing possibility to furnish a physical interpretation for the dark energy component is consider it as the energy stored on the vacuum state related to all fields that fill the universe. 
 
 However, if from experimental viewpoint of the accelerated expansion we have a paradigmatic model, in the side of the theoretical approach we do not have an identical picture.  As example, we have the well-known cosmological constant problem \cite{Weinberg}, that expose the apparent failure of the renormalization procedure to estimate a value for the vacuum energy in accord with the observational data \cite{Martin}.  Indeed, 	to harmonize both viewpoint, the scientific community  have considered a dynamical cosmological term $\Lambda (t)$.  
Recently, an emergent perspective of the gravity has been discuted in \cite{Pad1}, and the cosmological constant problem from the emergent gravity perspective can be found in \cite{Pad2}

Now, if we see the universe expansion and the gravitational collapse as different sides of the same coin (using other words: considering the process of collapse as the universal expansion with the inverted time), emerges  a natural question: What is the influence of the energy stored in the vacuum on the gravitational collapse process?  To answer this question and forecast the final fate of a collapsing star , the seminal study performed by Oppenheimer and Snyder, about the homogeneous spherically symmetric collapse \cite{Oppenheimer}, sets the concept of the trapped surface as fundamental for a clear understanding of the final stage for the  collapse process.  In this case we always get a black hole recovering the singularity, not contradicting the cosmic censorship conjecture (CCC). 

The cosmic censorship conjecture of Roger Penrose establishes that all spacetime singularities are always hidden behind an event horizon.  Using other words, naked singularities do not exist.
 However, breaking the spacetime homogeneity by introducing inhomogeneities, it is possible that naked singularities are formed \cite{Josh}, performing a counterexample for CCC. 

The importance of the CCC can be noted in the continuous collapse of a star which has exhausted its nuclear fuel, or in the black hole physics, to mention two immediate examples.  Presently, we do not have a satisfactory mathematical formulation of CCC available, but we have in scientific literature several counterexamples, besides of the above cited,  where the integration of the Einstein field equations result solutions which admit naked singularities. See for example the references \cite{Josh}-\cite{Gosh}.

On the other hand, the inclusion of higher dimensions is not only a playground for theoretical physicists. The gravitational field at high density matter and near of the singularity, where the quantum effects emerge, we can not discard the use of spacetime with different number of dimensions, than the usual four-dimensional spacetime. So, the Lanczos-Lovelock Lagrangean is the more natural generalization of the Einstein-Hilbert Lagrangean in arbitrary dimensions, and share several properties with Einstein's theory in four-dimensional spacetime, but in this work we will limit us at second option.

The original motivation to include additional dimensions in the spacetime that describe our universe was linked to an attempted to unify the electromagnetism and general relativity \cite{Kaluza}, \cite{Klein}, but with common-place in the theories where the extra dimension have a size comparable with the Planck length.  There are other scenarios where the inclusion of extra dimensions are not excluded.   We can cite as example, the experimental results using the inverse square law \cite{Adelberger}, and the ADS/CFT correspondence, that until we know was the last important theoretical contribution to the physics of fields, pointing for the not rule out of higher dimensional spacetimes \cite{Maldacena} \cite{Peterson}.  In this conjecture, we have the correspondence between field theories using isometric invariant fields in the five-dimensional anti-deSitter space, furnishing an additional theoretical option to define a quantum theory of gravitation \cite{Maia}.  To finish our brief list of examples to do not rule out the inclusion of extra dimensions we must cite the braneworld scenario, where additional space dimensions has been considered, gravity propagates in all dimensions, while matter is localized in the usual four-dimensional spacetime \cite{Randal}.

Considering a more robust theoretically viewpoint, naturally the spacetime have four dimensions, that is consistent with the Maxwell equations and Yang-Mills fields. However, as the Einstein theory of the gravitation do not have the same gauge structure, consequently it do not has the same  limitations in respect to the dimensionality \cite{Maia}.

Now, explicitly, in the case of the study of gravitational collapse in a spacetime with extra dimensions, we can cite the work of Goswami and Josh \cite{Goswami} that studied the effects of extra dimensions on a collapsing cloud, and in \cite{Rocha}, for a similar spacetime, appear solutions for a perfect fluid with state equation given by $P=\alpha \rho$, where we have the formation of the black holes with a null initial mass when the collapsing process has a continuous similarity. On the other hand, for the inhomogeneous collapsing process at higher dimensions, we have counterexamples of the hypothesis  of the CCC that have been studied in \cite{Banerjee} and \cite{Uggal}.

Finally, what is our purpose in this work, compared to what is stated above? We study the influence of the vacuum energy in the collapse process taking into account extra dimensions in the framework of the general relativity theory.  In a previous work, Campos \cite{Campos} and Campos $\&$ Lima \cite{Manuela} discussed this subject, responding in some sense about the influence of the vacuum energy and the formation of black holes. In our present study, we generalize this previous work including extra dimensions, where we have new questions: how works the competition between  the vacuum energy  in the stellar fluid and the inclusion of higher dimensions, in the final fate of the collapse process? 

Although the majority of the scientific community believe in the existence of a horizon  in the collapse of large amounts of mass, there are studies in the literature that advocate the opposite. For example, we can mention the work of Chapline and Barbieri \cite{Chapline} where the authors propose that due to the effect of a collective nucleon decay on the dynamics of the colapse process we do not have the emerging of a horizon, or even of a singularity.  However, we have a very strong point in favor of the formation of a horizon at work \cite {Bro}, where the authors show that recent observations of Sagittarius $A^\ast$ require the existence of a horizon.

So, we do not yet have a precise formulation of cosmic censorship conjecture, which makes it even interesting the discussion of the theoretical viewpoint; from the experimental viewpoint appear ways in the recent literature, that eventually can distinguish between naked singularities and black holes. Hence, to exemplify, the discrimination between black holes and naked singularities can be realized using the strong gravitational lensing and accretion disks \cite{Virbahadra}-\cite{Kovacs}. Subsequently, adopting the observational or theoretical point of view, the study of the ultimate fate of the gravitational collapse process is an important topic of current research. With this concern in mind, we finish this study show the common points and differences for the collapsed mass that is calculated using the Cahill-McVittie mass definition \cite{Cahill} from the definition introduced by Chatterjee-Bhui \cite{Bhui}. We discuss, also, about the active gravitational  mass \cite{Ellis} of the collapsed object, considering the mass definition of the Tolmann-Whittaker \cite{Tol}, \cite{Tol2}, \cite{Whi}, \cite{Mitra}.
\section{Basic equations for the higher dimensional gravitational collapse}
\label{sec:1}
In this section we discuss the basic equations of the gravitational collapse at higher dimensional spacetime where the material fluid interacts with the  vacuum  energy, in the framework of the Einstein's theory.

Let us consider the metric for $(N+2 )$-dimensional spacetime with spherical symmetry
\begin{equation}\label{1}
ds^2=dt^2-a(t)^2\left \lbrace dr^2 +r^2d\Omega ^2 _{N}\right \rbrace \, ,
\end{equation}
where
\begin{equation}
d\Omega ^2 _{N}=d\theta_1^2+...+\sin\theta_1^2 \sin\theta_2^2...\sin\theta_{N-1}d\theta_N^2
\end{equation}
 is the $N$-dimensional sphere. In spite of the spherical model be an ideal case, the main features of the collapse process can be seen, and important physical characteristics as the mass collapsed and formation of the apparent horizon can be studied.  Note that, we consider our "stellar fluid" in a higher dimension spacetime as simple generalization of the FRW solution, that naturally, can be matched to the exterior by Schwazschild-Thangerlini solution.

For the spacetime governed by the metric Eq.(\ref{1}), the Einstein filed equations can be written as:
\begin{equation}\label{2}
G^{\alpha \beta}=\kappa_{N+2}\left[ T^{\alpha \beta}+\frac{\Lambda}{\kappa _{N+2}}g^{\alpha \beta}\right]\, ,
\end{equation}
where $\kappa_{N+2}$ is the generalization for the Einstein gravitational constant for higher dimensions \cite{Barton}.  

Making use of the Bianchi identities results the conservation law $T^{\alpha \beta}_{;\beta}=0$, that is a valid equation if we have a constant cosmological term.  However, considering a time dependent cosmological term, we must to assume  some sort of interaction between the material of the fluid and the vacuum energy.

Then, to describe more adequately the above assumption, we consider  the energy-momentum tensor as a simple extension of the perfect fluid  for the usual four-dimensional case, namely
\begin{equation}\label{3}
T^{\alpha \beta}=(\rho +P)u^\alpha u^\beta -P g^{\alpha \beta} \, ,
\end{equation}
where $\rho$, $P$ and $u^\alpha $ are the total density that includes the material component plus the vacuum energy, the correspondent total pressure, and the (N+2)-dimensional velocity of a element of the fluid, respectively. 

Taking the divergence of the Eq. (\ref{2}) and projecting the results in the direction of the $(N+2)$-dimensional  velocity of a fluid element, one finds
\begin{equation}\label{4}
u_{\alpha}T^{\alpha \beta}_{;\beta}=-u_\beta\left(   \frac{\Lambda g^{\alpha \beta}}{\kappa_{N+2}}  \right)_{;\beta}\, ,
\end{equation}
that in our case reduces to:
\begin{equation}\label{5}
\dot{\rho_f}+(N+1)H(\rho_f+P_f)=-\dot{\rho_v}\, ,
\end{equation}
where $\dot{\rho_v}=\frac{\dot{\Lambda} (t)}{\kappa_{N+2}}$ is the time derivative of the vacuum energy density ($\rho_v$).  The subscript $f$ refers to fluid and the subscript $v$ to vacuum.

At this point of our work the equations of Einstein should have already appeared, that explicitly are given by:
\begin{eqnarray}\label{6}
\frac{N(N+1)}{2}\frac{\dot{a}^2}{a^2}&=&\kappa_{N+2} \rho_f +\Lambda  \, , \\
N\frac{\ddot{a}}{a}+\frac{N(N-1)}{2}\frac{\dot{a}^2}{a^2}&=&-\kappa_{N+2} P_f+\Lambda\, ,
\end{eqnarray}
that decays in the usual field equations for the four-dimensional case ($N=2$) \cite{Campos}.

In order to solve the Einstein field equations we need to specify the state equation and the function that govern the vacuum energy.  For the state equation we consider valid the usual expression $P_f = \omega \rho_f$, where $0 \leq \omega \leq 1$.  Besides, we assume that the variable vacuum energy interacts only with the dominant fluid component, and that this mixture determines the overall evolution of the collapse process.

In respect to the functional dependence for the vacuum energy  , let us see what appears in the literature. In the paper \cite{Overduin}, the authors present many different functional forms for $\Lambda (t)$.  However, as the list of suggestions for $ \Lambda (t) $ is not small, we will mention two important examples, that we think it is enough, to lead us to our "ansatz" for the vacuum dependence.  Hence, we have the work of Carvalho and Lima \cite{Carvalho} that consider $\Lambda \propto H^2$ based in dimensional arguments, and  based in a renormalization group approach, Shapiro and Sol\`a consider an identical dependence for $\Lambda(t)$ \cite{Shapiro}.

Consequently, in the absence of a fundamental theory that allow us to establish an expression for the $\Lambda$-term, we follow the cited authors above and adopt as "ansatz" for  the vacuum component  $\Lambda = \Lambda_0 + 3\beta H^2$, with $\beta$ constant.  However, as the goal here is quantify the influence of the vacuum energy in the last stages of the gravitational collapse at higher dimensional spacetimes,  we can neglect the bare cosmological constant ($\Lambda_0$).  With these considerations, we summarize  the Einstein field equations in the following differential equation:

\begin{equation}\label{7}
\frac{\ddot{a}}{a}+\delta \left(\frac{\dot{a}}{a}\right)^2=0 \, ,
\end{equation}
where $\delta = \frac{N-1}{2}+\frac{w\left(N+1\right)}{2}-\frac{3\beta\left(1+\omega\right)}{N}$\,. 
\begin{figure}
\label{fig.scale}
\includegraphics[scale=.6]{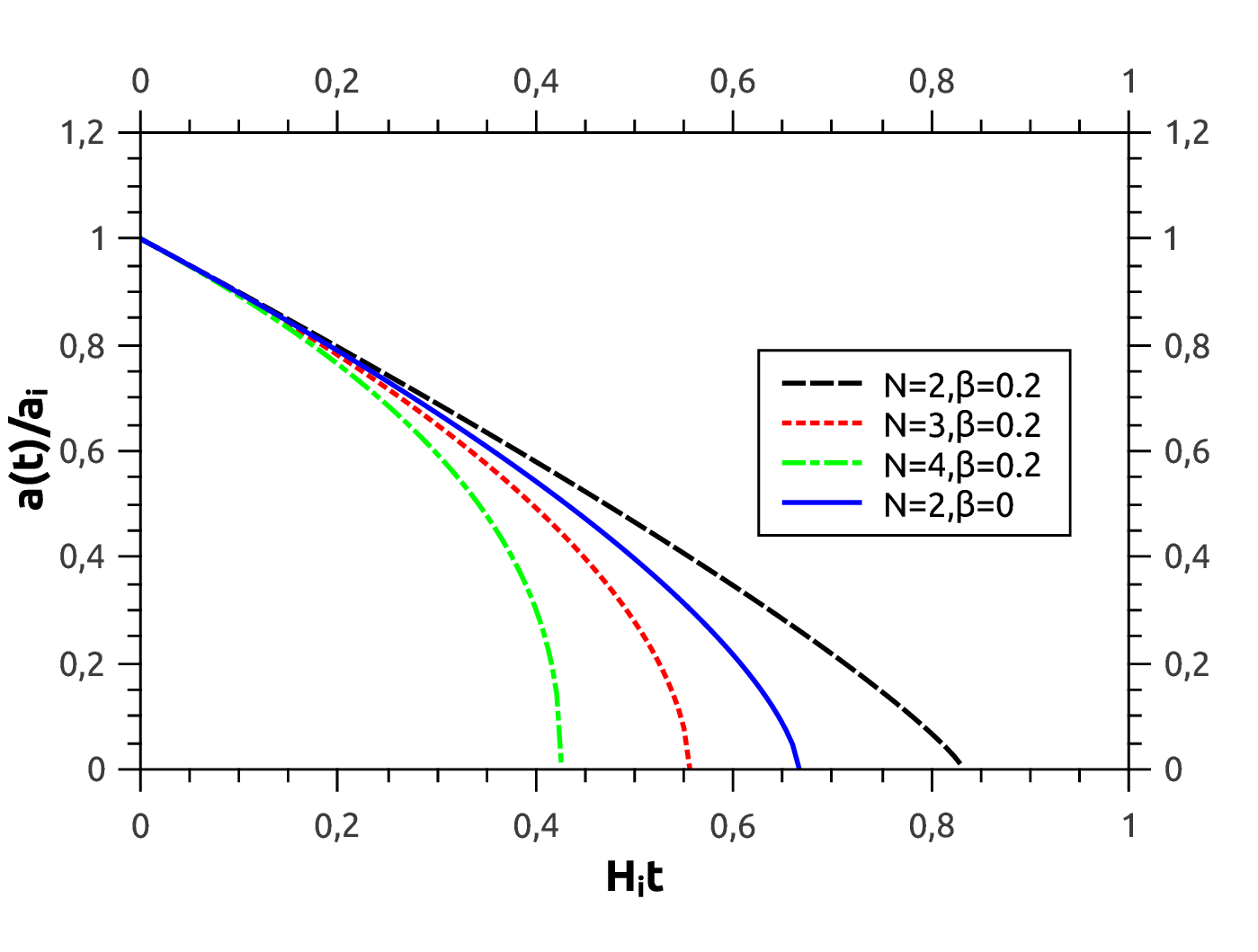}
\caption{Evolution of the scale factor, dust case ($\omega = 0$). We consider several values for the spacetime dimensions taking into account $\beta =0.2$, and $\beta=0$ for the four-dimensional spacetime. The inclusion of additional dimensions and the vacuum energy has opposite effects in respect to the necessary time to reaching the singularity.}
\end{figure}

The integration of Eq.(\ref{7}) is simple and results:
\begin{equation}\label{8}
a(t)=\left \lbrace  \left(1+\delta\right)\left( C_2+C_1 t\right)  \right \rbrace^{\frac{1}{1+\delta}}\, .
\end{equation}
To estimate the integration constants $C_1$ and $C_2$ we define the initial conditions $a(t=0)=a_i$ and $H(t=0)=-H_i$, that are compatible with the collapse process.  Then, the scale factor can be rewritten as
\begin{equation}\label{9}
a\left(t\right)=a_i\left \lbrace  1-(1+\delta)H_i t   \right \rbrace^{\frac{1}{1+\delta}}\,.
\end{equation}
It is worth commenting that apart from the physical choice of integration constants, the above solution reduces to the similar solutions derived by Campos and Lima (Eq.(12) in  \cite{Campos}) for the usual four-dimensional spacetime.  

Looking the evolution of the scale factor (see Fig.1), all solutions considered reach the singularity in a determined moment, that we write explicitly using Eq.(\ref{9}), obtaining
\begin{equation}\label{10}
t_c=\frac{1}{(1+\delta)H_i}\, .
\end{equation}
Resuming,  we write the scale factor and the Hubble function in terms of 
the collapse time ($t_c$), namely
\begin{eqnarray}\label{11}
a(t)&=&a_i\left( 1-\frac{t}{t_c}\right)^{\frac{1}{1+\delta}}\, , \\
H(t)&=& -\frac{H_i}{1-\frac{t}{t_c}}\, ,
\end{eqnarray}
that basically are the important quantities to proceed our study of the gravitational collapse.

 To complete our discussion of the singularity reaching we use the Raychauduri equation, that in mathematical terminology can be written as \cite{Ellis}, \cite{Alba}, \cite{Kar2} :
\begin{equation}\label{Ray}
\frac{d\theta}{d\lambda}+\frac{\theta ^2}{2}+\sigma ^2 - \omega ^2=-R_{\alpha \beta}u^\alpha u^\beta \, ,
\end{equation}
where $\theta$ is the expansion, $\sigma$ is the scalar shear, $\omega$ is the twist, $\lambda$ is an affine parameter and $R_{\alpha \beta}$ is the Ricci tensor.  We can write the Raychauduri equation as a second order linear differential equation, namely:
\begin{equation}
\frac{d^2 F}{d \lambda ^2}+\frac{1}{2}\left( R_{\alpha \beta}u^\alpha u^\beta +\sigma ^2 -\omega^2 \right)F=0 \, ,
\end{equation}
which is the equation of a harmonic oscillator with a time dependent frequency.

Hence, the convergence of the light rays occurs if
\begin{equation}
R_{\alpha \beta}u^\alpha u^\beta +\sigma ^2 -\omega ^2 \geq 0 \, ,
\end{equation}
that in our work assumes the form
$$\frac{5+N+3\delta}{6(1+\delta)}\leq 0\, ,$$
and we use the Sturm comparison \cite{Doms}.

To exemplify the above condition, consider the dust case and the four-dimensional spacetime ($N=2$). For this case we obtain $\beta<1$.  Substituting $\beta=1$ in the Eq.(\ref{7}) results a de Sitter solution, that as we know is not singular.

Naturally, nothing prevent us  from considering a different dependence for the vacuum energy density with inevitable modifications in all physical quantities and in the gravitational collapse process. So, in a recent paper, Lima et al. \cite{Limaa} realized a thermodynamic analysis of a cosmic scenario with accelerated expansion avoinding the inclusion of a dark energy component.  The authors defend a cosmic scenario with gravitationally particle production.  In a determined point of the paper the authors defend as appropriate to consider the rate of created particles as given by  $$ \Gamma = 3H \left[\frac{H}{H_I}\right]^n\, ,$$ where $n$ is a nonnegative constant parameter, and H is the usual Hubble function.  In the author's analysis is show that the generalized model adopted is possible to describe an universe free of a singularity.  In spite of the models with particle production and the inclusion of a $\Lambda$-term are physically quite differents, from the mathematical viewpoint, one mimics the other very well.

Hence, we can obtain an equation similar to the differential equation Eq.(8), 
$\dot{H}+2H^2[1-\frac{H}{H_I}]^n$, 
 in the cited work, making use of a parallel ansatz for the vacuum energy density, instead of use a source for particle production.  Apparently, although the cosmological viewpoint differs of the gravitational collapse process, the mathematical similarities are enough for possibility us to find identical conclusion, that is: we do not have a singularity formation.  This is a point that deserves a more detail study, and we will make timely.
\section{Apparent horizon and collapsing process}
We identify each element with a comoving observer, that allows us to think in a fictional region of constant radius (denoted by $r_\Sigma$) which separates the interior region from exterior.  In this context, the collapse process do not form crossing shell singularities \cite{Banerjee}, and the set of above properties makes the physical visualization of the apparent horizon surface easier, even, in our case, considering spacetimes with higher dimensions.
 
Hence, with the evolution of the collapsing process and the increase of the fluid density, the light that emerge from the interior region and that could be seen by an external observer becomes more misty, until the moment that the external observer loses the visual contact.  In this moment the gravity is so strong that the light remains trapped in the interior region, a horizon is formed covering the singularity, and we have the formation of a black hole.

The apparent horizon is a null surface of constant radius where we have future  null geodesics with a converging point in both sides of the surface  \cite{Hawking}.  The importance of emerging of an apparent horizon is due to the association with the final fate of the collapse process, and anything that hits  this null surface from the exterior, disappears from the field of view for any external observer.  Consequently, if we have the formation of the singularity after the emerging of the apparent horizon,  we have  a black hole formation, and in the opposite sense we have the formation of a naked singularity, in this case, contradicting the weak form for the conjecture of the cosmic  censorship. The weak form for the cosmic censorship conjecture states the impossibility of an observer at infinity see the singularity. To a more precise description of the weak form of the cosmic censorship, see for example \cite{Wald}  

The condition for  the  apparent horizon formation can be formulated  by the expression \cite{Hawking}
\begin{equation}\label{12}
R_{,\alpha}R_{,\beta}=(r\dot{a})^2-1=0 \, ,
\end{equation} where
$()_{,x}=\frac{\partial}{\partial x}$ and $R(t,r)=ra(t)$.

At the beginning of the collapse process we consider that the fluid medium is not trapped, and any surface inside the star follows: 
\begin{equation}\label{13}
R_{,\alpha}R_{,\beta}=\left[  r_\Sigma \dot{a} \right]^2 -1<0 \, ,
\end{equation}
which implies $0<R_iH_i < 1$.

Using the apparent horizon condition Eq.(\ref{12}) in the expression 
\begin{equation}\label{14}
\dot{R}=r\dot{a}=R_iH_i(1-\frac{t}{t_c})^{\frac{-\delta}{1+\delta}}\, ,
\end{equation}
we can find the ratio between the moment formation of the apparent horizon and the collapsing time, namely
\begin{equation}\label{15}
\frac{t_{AH}}{t_c}=1-(R_iH_i)^{\frac{1+\delta}{\delta}}\, ,
\end{equation}
that is a important expression to decide if the final fate of the collapse process is a black hole or a naked singularity.
\begin{figure}
\label{fig.tah}
\includegraphics[scale=0.6]{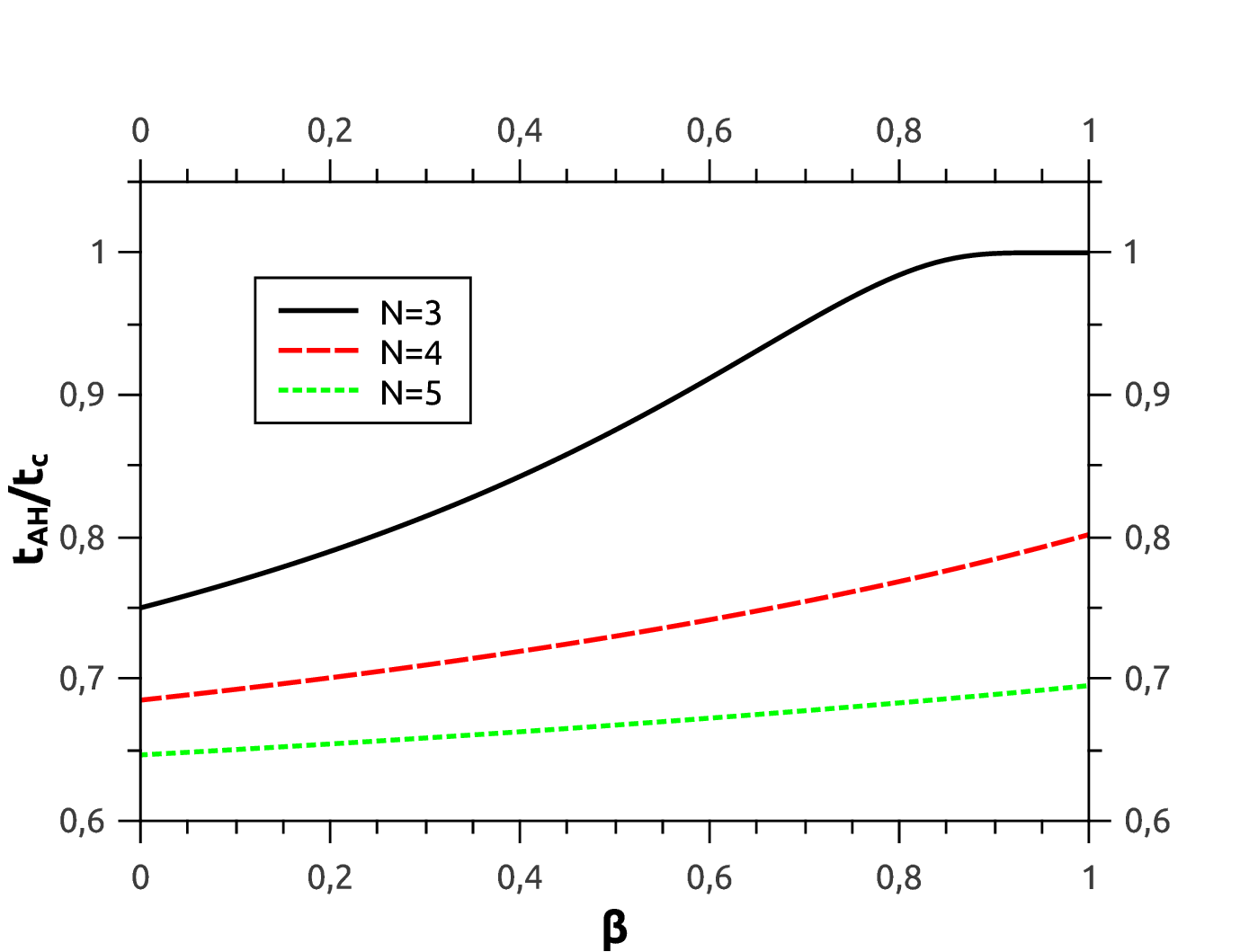}
\caption{Ratio between the apparent horizon moment formation and the collapse time versus the $\beta $-parameter, dust case. We consider several values for the spacetime dimensions and  $R_iH_i = 1/2 $.  With the increasing of the vacuum energy contribution, the apparent horizon moment and the collapse time are closer, but the increase of the number of dimensions has a opposite effect. }
\end{figure}

To guarantee the validity of the weak form for the  conjecture  of the cosmic censorship, and  obtain a black hole as final fate of the collapse process , we must to consider $0<\frac{t_{AH}}{t_c}<1$, that with help of Eq.(\ref{15}) reduces to
\begin{equation}\label{16}
\beta < \frac{N}{3(1+\omega)}\left \lbrace \frac{(N+1)(1+\omega)}{2} - 1 \right \rbrace \, .
\end{equation}
whose violation gives rise to a naked singularity.

From the cosmologival  viewpoint with dark sector, we have in the literature some constraints on the $\beta$-parameter.  For example, Birkel and Sarkar \cite{Drico} derived the limit $\beta < 0.13$ using the primordial nucleosynthesis in a model with decaying of the vacuum.  This upper limit was upgraded by Lima et al., founding $\beta \leq 0.16$ \cite{Batista}.  On the other hand, Basilakos \cite{Dinair} obtained $\beta \sim 0.004$ studing interacting dark energy models, with dark matter decaying.

Examining the limit for the $\beta$-parameter that point to different fates for the final of the collapsing fluid, or using different words, the value for the $\beta$-parameter that remain valid the cosmic censorship conjecture, we note that the constraints outlined in the literature are in accord with the condition given by Eq.(\ref{16}), in spite of our case (gravitational collapse process) is permitted have higher values for $\beta$, remembering that in our study the number of dimensions has a direct influence in the upper limit for the parameter associated with the vacuum decaying.

To clarify the influence of the spacetime dimensions and the vacuum energy in the collapse process, we display in the Fig.(2) the dependence of the ratio between the moment formation of the apparent horizon  and the collapse time versus the $\beta$-parameter, considering different number of dimensions.  In the range  $\frac{t_{AH}}{t_c} < 1$ we have only dressed singularities , since we do not was enough time to form the  apparent horizon, before the reaching of  singularity.

Although we  do not have a closed theory about the dynamics of horizons; we already have in the literature  some works using the observational viewpoint, that discuss possible experiments for differentiate among a naked singularity and a black hole.  In the paper by Joshi and Malafarina \cite{Malafarina}, interesting aspects about gravitational collapse and spacetime singularities are discussed: as the possibility to test the cosmic censorship conjecture; the equilibrium configuration describing an extended compact object obtained from gravitational collapse; and distinguishing black holes of naked singularities.
Nevertheless, observations involving strong gravitational lensing and accretion disks are able to discriminate black holes of naked singularities \cite{Virbahadra}, \cite{Sahu}, \cite{Kovacs}.  Consequently, these feasible experiments  point to the  collapsed mass as a fundamental physical quantity , able to perform the differentiation among distinct options for the final fate of the collapsing objects.  Albeit, in the \cite{PJoshi} the authors investigated whether the models for naked singularities  can be observationally distinguished of black holes with identical mass.

Since the collapsed mass is an important concept in the process of collapse, we calculate the mass for the collapsed object considering two distinct concepts: the Cahill McVittie definition for higher dimensional spacetime, and the suggestion of Chaterjee and Bhui \cite{Bhui}.  Taking into account the first option, the collapsed mass resulting is given by
\begin{equation}\label{eq.mass}
m(t,r)=\frac{R\dot{R}^2}{2}=\frac{R_i^3H_i^2}{2}\left(1-\frac{t}{t_c}\right)^{\frac{1-2\delta}{1+\delta}}\, .
\end{equation}
\begin{figure}\label{fig.mass}
\includegraphics[scale=.6]{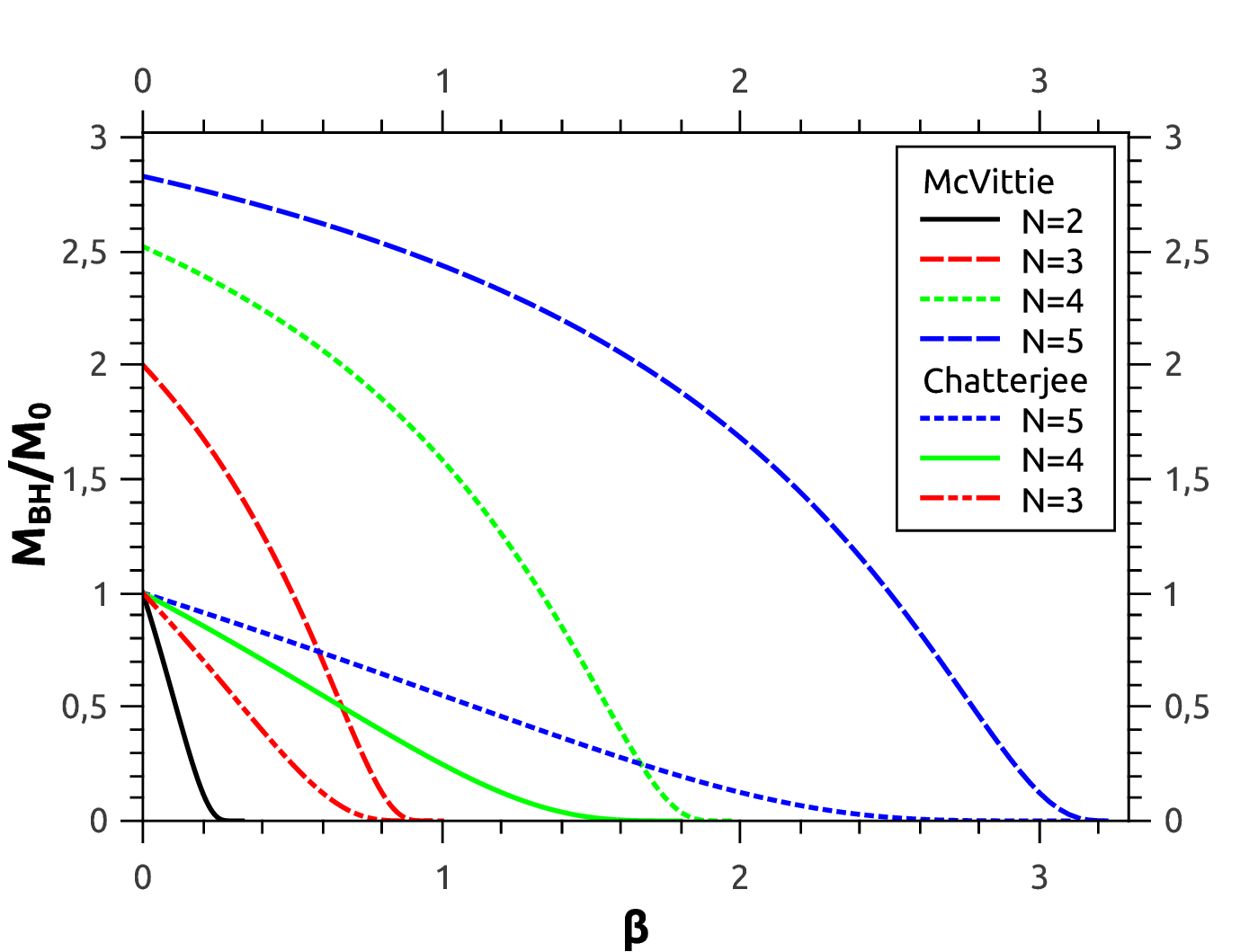}
\caption{Dimensionless mass of the formed black hole, dust case. We display the dimensionless black hole mass versus the $\beta$-parameter taking into account the usual Cahill-McVittie definition for the collapsed mass.  Also appear in the graph the Chatterjee-Bhui suggestion for the collapsed mass, and we can note that in this new approach we obtain black holes with less mass than in Cahill-McVittie case. For the four-dimensional case ($N=2$) both definitions are identical.}
\end{figure}

Initially the fluid medium is not trapped, and after the apparent horizon emerging, immediately, all the mass collapse inside it.  Therefore, the mass of the black hole formed can be calculated using the expression for the collapsed mass at the apparent horizon moment, namely:
\begin{equation}\label{bh1}
M_{BH}=m(r,t=t_{AH})=\frac{R_i^3H_i^2}{2}\left(1-\frac{t_{AH}}{t_c}\right)^{\frac{1-2\delta}{1+\delta}}\, ,
\end{equation}
that, with help of  Eq.(\ref{15}), can be written as
\begin{equation}
\frac{M_{BH}}{M_0}={(R_iH_i)}^{\frac{1-2\delta}{\delta}}\, ,
\end{equation}
where $M_0=\frac{1}{2}R_i^3H_i^2$ is the mass of the dust pure fluid.
In this fashion, the black hole mass calculated is, in the truth, an upper limit of the mass  that fallen within the surface that defines the apparent horizon \cite{Krasinski}. 

Looking for an alternative to calculating of collapsed mass, some time ago Chatterjee and Bhui, built an alternative expression for higher dimensions, following similar steps of the  work of the Cahill and Mc Vittie.  Hence, applying the mass definition of Chatterjee and Bhui, results for our case:
\begin{equation}
\tilde{m}=\frac{N-1}{2}R^{N+1}\frac{\dot{R}^2}{R^2}\, ,
\end{equation}
that furnishes for the mass of the black hole formed 
\begin{equation}\label{bh2}
\frac{\tilde{M}_{BH}}{M_0}\propto {(R_iH_i)}^{\frac{N-1-2\delta}{\delta}}\, . 
\end{equation}
Although, the black hole mass in both cases are identical for the usual four-dimensional case, at higher dimensions the  suggestion of Chatterjee favoured the formation of black holes with less mass than the original definition of Cahill and McVittie,  see Fig.(\ref{fig.mass}).

In the graph for the black hole mass versus the vacuum parameter appears the possibility of obtain a black hole with null mass. In our work, this characteristic is not due to the emission of thermal radiation known as Hawking radiation; or  mass ejection.  This is possible in our model due to the contribution of the vacuum component to the material fluid.  Let us detail some points about this characteristic, rewritten the field equation given by Eq.(7) as
\begin{equation}\label{po1}
\frac{a \dot{a}^2}{2}=\frac{\kappa _{N+2}}{N(N+1)}\rho a^3\, ,
\end{equation}
that is the well-known expression of the Misner-Sharp-Ernandez mass \cite{Ernandez}, \cite{Misner}.  Comparing with the mass definition of Cahil and McVittie (given by the Eq.(\ref{eq.mass})), we find
\begin{equation}\label{null-mass}
m(t,r) = \frac{\kappa _{N+2}}{N(N+1)} \rho  R^3\ \, .
\end{equation}
In spite of this comparison, that clarifies the mass definition, 
is still not very clear how the collapsed mass can be null.

However, using again the Einstein field equations we can write the suggestive equation
\begin{equation}\label{acc}
\frac{\ddot a}{a}=-\frac{\kappa_{N+2}}{N(N+1)}\left \lbrace  (N+1)P +(N-1)\rho   \right \rbrace \, .
\end{equation}
We think that is premature  to link directly the collapsed mass with the quantity $\ddot{a}$, considering valid in this context a version of the Newton second law. Despite this, for $\beta = \frac{N}{3(1+\omega)}\left \lbrace \frac{(N+1)(1+\omega)}{2} - 1 \right \rbrace$, the Eq.(\ref{15}) results $t_{AH}=t_c$, Eq.(\ref{bh2})results $M_{BH}=0$, and $\frac{\ddot a}{a}=0$.
    
    At first seems that these quantities have some link between them, but note that the Cahill-McVittie mass definition do not includes what which would an active definition \cite{Ellis} for the collapsed mass, or with other words, the inclusion of the pressure is not contemplated in the Cahill-McVittie mass definition.  On the other hand, the acceleration of the element fluid, given by Eq.(\ref{acc}), has the pressure contribution, and the right side of Eq.(\ref{acc}) is null for $P=-\frac{(N-1)}{(N+1)}\rho$ (considering the four-dimensional case results $P=-\rho/3$).  To avoid that both sides of Eq.(\ref{acc}) should be nulls independently, we can rewrite the state equation by
\begin{equation}
P=\frac{N(1-N)+6\beta}{N(1+N)-6\beta}\rho \, ,
\end{equation}
where we substitute the $\omega$-parameter of the state equation, using the condition given by Eq.(\ref{acc}).  To exemplify and clarify, let us consider the dust case. Consequently, we have $$ \frac{N(1-N)+6\beta}{N(1+N)-6\beta}=0$$, and substituting $N=5$ we find $\beta=20/6$, in accord with appear in the graph for the black hole mass in the Fig.(3) ($M_{BH}/M_0 = 0$ for $\beta = 3.33$).
\section{Final Comments}
In this work we discuss a version with an arbitrary number of dimensions for the  influence of the vacuum energy in the gravitational collapse.  We consider the spacetime spherically symmetric with a finite radius filled by a homogeneous and isotropic fluid plus an interacting vacuum energy density.  Besides, we neglected the curvature effects and  considered the vacuum term given by $\Lambda = 3\beta H^2$, where we  disregard also the bare cosmological constant.

In our study we conclude that the vacuum energy and the inclusion of additional dimensions compete at different directions in the collapse process.  While the vacuum energy hampers the collapse process, the addition of extra dimensions favour the  reaching of the singularity (see Fig.(1)).  Besides, the collapse time prone to for the infinity when $\beta = \frac{N(N+1)}{6}$ (dust case), obtaining in this case a not singular model.
Note that for $N=2$ we obtain $\beta=1$, and substituting this value for $\beta$ in the Eq.(\ref{7}) results in a pure de Sitter vacuum solution, namely $a(t)\propto \exp{C_1 t}$ ($C_1$ is a integration constant), that is not a singular solution.

In respect to the formation of the apparent horizon formation we obtain a specific value for the $\beta$-parameter, namely 
$$\beta=\frac{N}{3(1+\omega)}\left\lbrace \frac{(N+1)(1+\omega)}{2} -1  \right\rbrace \, .$$
Considering values for the $\beta$-parameter greater than the value above, the reaching of the singularity occurs before the necessary time to the formation of the apparent horizon.  On the other hand, for smaller value for the $\beta$-parameter than the value above, we have the opposite happening.  Resuming, in the first case the naked singularity is favoured, while in the second case we have the formation of a black hole.

In the final of the last section we discuss the collapsed mass of the black hole formed using the definition of the Cahill-McVittie and the definition of the Chaterjee-Bhui.  In spite of both definitions are identical for the four-dimensional case ($N=2$), for $N>2$ the Cahill-McVittie definition furnishes black holes with higher mass. 
Still on the mass of black holes, there is another point that worth mention, we note in the Fig.(3) that is possible to obtain black holes with null mass, and this occurs for the same value of the $\beta$-parameter that separates different fates for the collapse process.  In our case the null mass for black holes is possible due to the contribution of the vacuum energy, but although we discuss this possibility using Eq.(\ref{po1}) and Eq.(\ref{acc}), furnishing some indications, we do not have yet a robust definition, that clarifies the mechanism that can produce a collapsed object with null mass.

In respect to the active feature of the gravitational mass, the author in \cite{Mitra} discuss  the local viewpoint of the Tolmann-Whittaker mass definition (TW).  Hence, in our work this definition assumes the form$$ M=\int _0 ^R [\rho+(N+1)P]\sqrt{g_{00}}dV$$, where $dV$ refers to the proper volume element.  Taking into account the integrand in above expression, the field equation Eq.(7), and the state equation $P=\omega \rho$, we obtain a null mass for the colapsed object when $$\beta= \frac{N}{6}\left[\frac{1+\omega(1+N)}{1+\omega}\right]\, .$$  This value for the $\beta$-parameter differ form the previously estimated from equation Eq.(27), namely: $$\beta=\frac{N}{6}\left[\frac{1+\omega(N+1)}{1+\omega}+\frac{N-2}{1+\omega}\right]\, .$$  However, for the usual four-dimensional ($N=2$) case, both expressions are equals, resulting $\beta=\frac{1}{3}$ for the dust case.

Currently, we are looking for alternatives to explain the difference that we found in the $\beta$ parameter, and briefly put our results on the subject.

\end{document}